\documentclass[12pt]{elsarticle}
\usepackage{amssymb}
\usepackage{amsmath}
\usepackage{bm}
\usepackage{amsthm}
\theoremstyle{definition}

\newtheorem{thm}{Theorem}
\usepackage{color}
\usepackage{empheq}
\newcommand{\red}{\color{black}}

\newcommand{\recolor}{\color{black}}
\journal{Physica D: Nonlinear Phenomena}
\begin{document}
\begin{frontmatter}

\title{Construction of cubic nonlinear lattice free from umklapp processes}
\author{Hiroki Ono}
\author{Yusuke Doi\corref{cor1}}
\ead{doi@mech.eng.osaka-u.ac.jp}
\cortext[cor1]{Corresponding author}
\author{Akihiro Nakatani}
\affiliation{organization={Division of Mechanical Engineering, Graduate School of Engineering\\ The University of Osaka},
            addressline={2-1 Yamadaoka}, 
            city={Suita},
            postcode={565-0871}, 
            state={Osaka},
            country={Japan}}

\begin{abstract}
We propose a novel type of umklapp-free lattice (UFL), where umklapp processes are completely absent. 
The proposed UFL incorporates cubic long-range nonlinearity, a feature not addressed in previous studies.
In this paper, we derive an analytical expression for the cubic nonlinear coupling constants by imposing  mathematical conditions such that the nonlinear coupling strength between particle pairs decays inversely with their separation distance.
The absence of umklapp processes in the proposed lattice is confirmed through numerical  comparisons with the Fermi-Pasta-Ulam-Tsingou (FPUT) lattice.
Furthermore, molecular dynamics simulations are performed to investigate the thermal conductivity of the proposed lattice in the non-equilibrium steady state.
Compared to the original FPUT lattice, the proposed UFL is closer to ballistic transport. 
Our results demonstrate that the umklapp processes induced by cubic nonlinearity are suppressed in the proposed UFL.
Moreover, compared to the UFL with only quartic nonlinearity, truncation of long-range interactions plays a significant role in the proposed lattice.
\end{abstract}
\begin{keyword}
phonon thermal transport, nonlinearity, symmetry
\end{keyword}
\end{frontmatter}

\section{Introduction}\label{sec1}
Nonlinear lattice models have been developed to study vibrations, wave propagation, and energy transport in crystalline solids from the perspective of nonlinear dynamics~\cite{kittel}.
Fermi-Pasta-Ulam-Tsingou (FPUT) lattice has been constructed to investigate the relationship between nonlinear interactions in particles and ergodicity~\cite{FPU}.
The study has attracted significant attention for demonstrating periodicity in temporal evolution as evidence of Poincar\'e's recurrence paradox through numerical simulations, showing that nonlinearity did not necessarily lead to thermal equilibrium or an equal distribution of energy.
Toda~\cite{toda1967_1} constructed another famous lattice with an exponential-type nonlinear particle interaction.
The lattice could be used to construct an integrable system and contribute to the development of soliton theory owing to its potential for analytical studies~\cite{toda1967_2}.
The temporal evolution of the displacement field of solids in a continuum can be transformed from Fourier's representation to wave-mode system dynamics. 
Similarly, waves in lattice systems can be understood in terms of vibration mode dynamics as phonons.
Peierls~\cite{peierls1929} identified the process in which the law of conservation of momentum did not hold in the interaction between phonons during umklapp processes. %
The process was defined as the origin of thermal resistance, arising from disturbances in energy transport.
Umklapp processes do not occur in linear lattices, where particle interactions are governed by a harmonic potential function. In such systems, no energy exchange takes place among harmonic vibration modes via linear interactions, implying that the energy of each phonon remains unchanged.
Energy transport realized by wave propagation is ballistic in linear lattices.
Therefore, the transition from vibration to heat and the resulting heat conduction, essentially originates from the nonlinearity of the lattice, where the interaction between phonons is usually significant and the superposition principle is violated~\cite{toda_nwas}.
Regarding problems in the energy transport of lattice systems, several studies have been conducted to clarify the relationship between the energy transport and phonon interactions~\cite{lepri2003,dhar2008,lepri2023_rev}.
Besides energy transport with nonlinearity, long-range interactions are also attached to many interests from experimental systems, such as magnetic lattices~\cite{Molerón_2019} and quantum systems~\cite{RevModPhys.95.035002}. 
Nonlinear lattices with only nearest-neighbor interactions in homogeneous particle systems were first investigated in earlier studies, such as \red FPUT-\(\beta\)~\cite{lepri1998,dematteis2020}, FPUT-\(\alpha\)~\cite{lepri2000}, and rotor model~\cite{PhysRevLett.84.2144,PhysRevLett.84.2381,Iubini_2016}. \recolor
One extension considers spatial heterogeneity, such as impurity particles and diatomic Toda lattice~\cite{toda1979,hatano1998}.
Another extension considers the long-range interactions of particles~\cite{livi2020,benenti2020,Andreucci2025}.
Long-range FPUT--$\beta$~\cite{bagchi2017thermal, bagchi2021,PhysRevE.106.014135,Christodoulidi_2014,xiong2022}, and other lattices~\cite{yoshimura2019nolta,yoshimura2020nolta,iubini2018,olivares2016,PhysRevE.108.024115} have been utilized to understand the nonlinear dynamics.
For example, pairwise interaction symmetric lattice (PISL)~\cite{doi2016,doi2022} supports the smooth mobility of discrete breathers~\cite{st1988,flach2008,Yoshimura2015} exhibiting higher thermal conductivity than the original FPUT-$\alpha$ and FPUT-$\beta$ lattices~\cite{yoshimura2019nolta,yoshimura2020nolta}.
Bagchi investigated the thermal conductivities of long-range FPUT lattices~\cite{bagchi2017thermal,bagchi2021}.
They determined the coupling constant that realized the maximum thermal conductivity.
Wang \textit{et al}. investigated the thermal conductivity of long-range FPUT lattice and discussed the role of discrete breather in thermal transport~\cite{PhysRevE.111.054102}.
However, the mechanism of thermal resistance has not been fully understood from the perspective of phonon interactions.
Yoshimura \it et al. \rm recently constructed a nonlinear lattice named umklapp free lattice (UFL)~\cite{yoshimura2022}. 
The lattice was constructed to satisfy the \(\mathcal{N}\mathcal{P}\) symmetry condition, which represents invariance against a specific mapping~\cite{yoshimura2023}. 
Mathematical proofs validated the absence of umklapp processes in the UFL system dynamics.
Furthermore, numerical simulations were performed to verify the absence of umklapp processes, which were closely related to ballistic energy transport.
Only the quartic nonlinear terms were considered in the interparticle potential function of the UFL.
Expanding the UFL to other orders of nonlinearity is crucial to generalize the UFL for understanding the effect of umklapp processes on thermal resistance.
For example, cubic nonlinearity is the lowest nonlinearity and important in understanding material properties considering nonlinear dynamics, as harmonicity only provides linear theories.
In the low-temperature regime, the possibility of thermal conductivity convergence due to cubic nonlinearity has been reported~\cite{chen2016}.
%
%

%
The aim of this study is to construct a UFL with cubic nonlinearity in the interparticle potential function.
The construction procedure is formulated from the perspective of lattice symmetry, ensuring complete suppression of umklapp processes.
The absence of umklapp processes is verified through numerical simulations of phonon excitations.
Furthermore, ballistic energy transport in the proposed UFL is confirmed via non-equilibrium molecular dynamics simulations, and the relationship between lattice symmetry and thermal transport is discussed. 
The remainder of this paper is organized as follows.
Sec. II introduces variable transformation and defines symmetry.
In Sec. III, the UFL is constructed by considering the symmetry in a cubic nonlinear potential.
The numerical evidence that umklapp processes vanish in the proposed UFL, is described in Sec. IV.
Ballistic energy transport in the proposed UFL is discussed in Sec. V by performing the non-equilibrium molecular dynamics simulation.
Sec. VI concludes the study.
\section{\label{sec:level2} Definition of symmetry in nonlinear lattices}
In order to construct the UFL, variable translation and potential mapping are introduced following existing studies \cite{doi2022,yoshimura2023}. 
Considering a one-dimensional (1D) nonlinear lattice defined by the Hamiltonian
\begin{equation}
    \mathcal{H}(\bm{q}, \bm{p}) = \sum_{n=1}^{N} \frac{1}{2}p_n^2 
    + \Phi (q_1,q_2,\ldots, q_N) \label{eq2.1.1},
\end{equation}
where \(\bm{q} = (q_1, q_2, \ldots, q_N)\) and \(\bm{p} = (p_1, p_2, \cdots, p_N)\), \(q_n, p_n \in \mathbb{R}\) represent displacement from the equilibrium point and momentum of the \(n\)th particle, respectively, \(N\) is an even number and corresponds to the number of particles, and \(\Phi: \mathbb{R}^N \rightarrow \mathbb{R}\) is a \(C^2\) function representing potential.
Assuming that the system has relations \(q_{N+n} = q_n, p_{N+n} = p_n, ~\ n=1,2,\ldots, \red L \recolor\) as periodic boundary conditions, where \(\red L \recolor\) is the truncation length corresponding to the range of pairwise interactions.
Introducing a complex normal mode coordinate \(U_m, V_m \in \mathbb{C}, m = -N_{\mathrm{h}}, -N_{\mathrm{h}}+1, \ldots, N_{\mathrm{h}} +1\) via the variable transformation defined by
\begin{align}
    q_n &= \frac{1}{\sqrt{N}} \sum_{m = -N_{\mathrm{h}}}^{N_{\mathrm{h}} + 1} U_m \exp \left(
        -\mathrm{i} \frac{2\pi n}{N}m
    \right), ~\ n = 1,2,\ldots, N, \label{eq2.1.2} \\
    p_n &= \frac{1}{\sqrt{N}} \sum_{m = -N_{\mathrm{h}}}^{N_{\mathrm{h}} + 1} V_m \exp \left(
        -\mathrm{i} \frac{2\pi n}{N}m
    \right), ~\ n = 1,2,\ldots, N, \label{eq:Vm}
\end{align}
where \(N_{\mathrm{h}} = N/2 - 1\) and \(\mathrm{i}\) is  the imaginary number unit.
\(V_m\) is defined as a derivative of \(U_m\) with respect to time.
Substituting Eq.~\eqref{eq2.1.2} and Eq.~\eqref{eq:Vm} into Eq.~\eqref{eq2.1.1}, the Hamiltonian can be rewritten in terms of $U_m$ and $V_m$ as
\begin{equation}
    \mathcal{H}(\bm{U}, \mathbf{V}, U_{N/2}, V_{N/2}) = \sum_{m=-N_{\mathrm{h}}}^{N_{\mathrm{h}}}  \frac{1}{2} V_m V_{-m} + \frac{1}{2}V_{N/2}^2
    + \Phi (\bm{U}, U_{N/2}), \label{eq2.1.4}
\end{equation}
where \(\bm{U} = (U_{-N_{\mathrm{h}}}, U_{-N_{\mathrm{h}}+1}, \ldots, U_{N_{\mathrm{h}}})\) and \(\bm{V} = (V_{-N_{\mathrm{h}}}, V_{-N_{\mathrm{h}}+1}, \ldots, V_{N_{\mathrm{h}}})\).
The potential $\Phi (\bm{U}, U_{N/2})$ can be divided into a $U_{N/2}$-independent component $\Phi_0(\bm{U})$ and a $U_{N/2}$-dependent component $\mathcal{G}(\bm{U}, U_{N/2})$.
Further, the following map \(\mathcal{T}_{\lambda}: \mathbb{C}^{N-1} \rightarrow \mathbb{C}^{N-1}\) is introduced: 
\begin{equation}
    \mathcal{T}_{\lambda}: U_m \rightarrow U_m \exp (-\mathrm{i}m\lambda),
   ~\ m = -N_{\mathrm{h}}, -N_{\mathrm{h}}+1, \ldots, N_{\mathrm{h}}
    \label{eq2.2.1}
\end{equation}
where \(\lambda\) is an arbitrary number.
We define the symmetry in the nonlinear lattice~\eqref{eq2.1.1} as the invariance of potential \(\Phi (\bm{U}, U_{N/2})\) in Eq.~\eqref{eq2.2.2} under the transformation \(\mathcal{T}_{\lambda}\) for any $\lambda\in \mathbb{R}$.
The $U_{N/2}$-independent component \(\Phi_0(\bm{U})\) can be divided into two parts as
\begin{equation}
    \Phi_0(\bm{U}) = \Phi_{\mathrm{s}} (\bm{U}) + \Phi_{\mathrm{a}}(\bm{U}). \label{eq2.2.2a}
\end{equation}
\(\Phi_{\mathrm{s}}(\bm{U})\) is defined as the symmetric part and it is invariant for mapping \(\mathcal{T}_{\lambda}, \forall \lambda \in \mathbb{R}\).
The relationship $\Phi_{\mathrm{s}}(\mathcal{T}_\lambda \bm{U})=\Phi_{\mathrm{s}}(\bm{U})$ holds for any $\bm{U}\in\mathbb{C}^{N-1}$ or $\lambda\in\mathbb{R}$.
$\Phi_{\mathrm{a}}(\bm{U})=\Phi_0(\bm{U})-\Phi_{\mathrm{s}}(\bm{U})$ is defined as the asymmetric part.
Then, \(\Phi (\bm{U}, U_{N/2})\) can be divided into three parts as
\begin{equation}
    \Phi (\bm{U},U_{N/2}) = \Phi_{\mathrm{s}} (\bm{U})+ \Phi_{\mathrm{a}} (\bm{U})+ \mathcal{G} (\bm{U},U_{N/2}). \label{eq2.2.2}
\end{equation}
The lattice is symmetrical when the following condition is satisfied:
\begin{equation}
    \Phi_{\mathrm{a}} (\bm{U}) + \mathcal{G}(\bm{U}, U_{N/2})=0. \label{eq2.2.3a}
\end{equation}
As will be discussed in the following section, symmetricity and asymmetricity correspond to normal and umklapp processes of energy transport, respectively.
If Eq.~\eqref{eq2.2.3a} is satisfied, then the nonlinear lattice model~\eqref{eq2.1.1} is defined as the UFL.
\section{\label{sec:level3}Construction of a cubic umklapp-free lattice}
The UFL with cubic nonlinearity is constructed by considering symmetricity in the potential.
We focus on a 1D lattice defined by the Hamiltonian
\begin{equation}
    H(\bm{q}, \bm{p}) = \sum_{n=1}^N \frac{1}{2} p_n^2 + \sum_{n=1}^N \frac{1}{2} \left(
        q_{n+1} - q_n
    \right)^2 + \alpha\sum_{n=1}^N \red\sum_{l = 1}^{L}\recolor \frac{1}{3}\red a_l \recolor \left(
        q_{n+\red l \recolor} - q_n
    \right)^3, \label{eq3.1.1}
\end{equation}
where \(\alpha \in \mathbb{R}\) denotes the factor of the cubic nonlinear potential. 
The factors \(\red a_l \recolor \in \mathbb{R}, \red l \recolor = 1,2, \ldots, \red L \recolor\) are constants that represent the coupling strength between the \(r\)th neighboring particles.
\begin{figure}[ht]
\centering
\includegraphics[width = 0.75\textwidth]{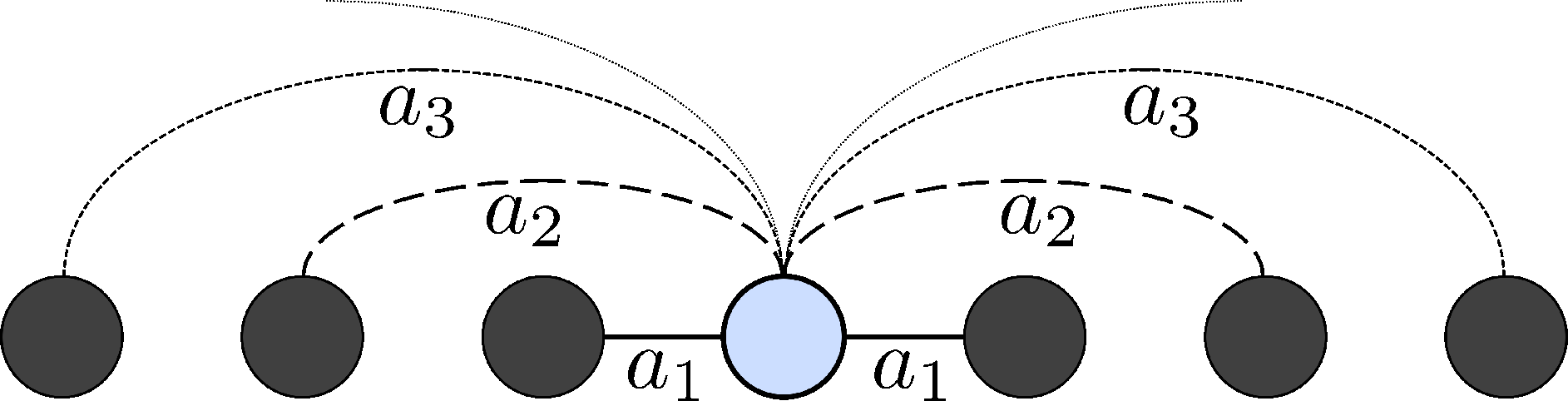}
\caption{The nonlinear lattice model for considering the UFL. The cubic long-range interaction is included. Only the connections related to the blue particles are shown.}
\label{fig:periodic}
\end{figure}
The cubic long-range interaction is considered as shown in Fig.~\ref{fig:periodic}.
Periodic boundary conditions are applied to the system, and the range of \(\red L \recolor\) is restricted to \(\red L \recolor \leq N/2\).
\subsection{Derivation of the condition of the symmetry}
As a first step to construct the cubic UFL, the condition of the symmetry is derived from Eq.~\eqref{eq3.1.1}.
Substituting Eq.~\eqref{eq2.1.2} into Eq.~\eqref{eq3.1.1}, the Hamiltonian is represented in terms of \(\bm{U}\), \(\bm{V}\), \(U_{N/2}\) and \(V_{N/2}\) as
\begin{align}
    &H(\bm{U},\bm{V}, U_{N/2}, V_{N/2}) \notag \\
     &= \sum_{m = -N_{\mathrm{h}}}^{N_{\mathrm{h}}} \frac{1}{2} \left(
        V_mV_{-m} + \red 4\sin^2 \recolor \frac{\pi m}{N} U_m U_{-m}
    \right)  \notag \\  &+ \frac{\alpha}{3\sqrt{N}}
     \sum_{i = -N_{\mathrm{h}}}^{N_{\mathrm{h}}} \sum_{j = -N_{\mathrm{h}}}^{N_{\mathrm{h}}} \sum_{k = -N_{\mathrm{h}}}^{N_{\mathrm{h}}} U_iU_jU_k G(\red L \recolor,i,j,k) \Delta(i+j+k) \notag \\
     &+ \frac{\alpha}{3\sqrt{N}}
     \sum_{i = -N_{\mathrm{h}}}^{N_{\mathrm{h}}} \sum_{j = -N_{\mathrm{h}}}^{N_{\mathrm{h}}} \sum_{k = -N_{\mathrm{h}}}^{N_{\mathrm{h}}} U_iU_jU_k  G(\red L \recolor,i,j,k) \Delta(i+j+k \pm N)\notag \\
     &+ \frac{V_{N/2}^2}{2} +  \mathcal{G}(\bm{U}, U_{N/2}), \label{eq3.1.5} 
\end{align}
where 
\begin{equation}
    G(\red L \recolor,i,j,k) = \red \sum_{l =1}^{L} \recolor \red a_l \recolor g(i) g(j) g(k),  
\end{equation}
\begin{equation}
    g(m) = \exp\left(-\mathrm{i} \frac{2\pi m}{N}\right) - 1, ~\ m = -N_{\mathrm{h}}, -N_{\mathrm{h}}+1, \ldots, N_{\mathrm{h}},
\end{equation}
\begin{empheq}[left={\Delta(d)=\empheqlbrace}]{alignat=2}
    1 & \quad (d = 0) \\
    0 & \quad (\mathrm{otherwise}).
\end{empheq}
The second and third terms and \(\mathcal{G}(\bm{U}, U_{N/2})\) on the right-hand side (RHS) induce three phonon processes.
Assuming that \(U_{N/2} \equiv 0\), \(V_{N/2}^2/2\) and \(\mathcal{G}(\bm{U}, U_{N/2})\) vanish.
Then, the condition of the symmetry in the nonlinear lattice~\eqref{eq3.1.1} is given as 
\begin{equation}
    \red \sum_{l = 1}^{L} \recolor (-1)^{\red l \recolor} \red a_l \recolor \sin \frac{\red l \recolor i \pi}{N} 
    \sin \frac{\red l \recolor j \pi}{N} \sin \frac{\red l \recolor k \pi}{N} = 0,  \label{eq3.1.2}
\end{equation}
where \(i, j, k \in \mathbb{N}\) are any numbers that satisfy
\begin{equation}
    2 \leq i \leq j \leq k \leq N_{\mathrm{h}}, ~\ i+j+k=N \label{eq3.1.3}
\end{equation}
(Appendix A presents derivation details).
It is worth discussing the physical interpretation of Eq.~\eqref{eq3.1.2}. 
The equation of motion in terms of the complex normal mode coordinates $U_m$, derived from the Hamiltonian \eqref{eq3.1.5}, is expressed as
\begin{align}
    &\ddot{U}_m + 4\red \sin^2 \recolor \frac{\pi m}{N} U_m \notag \\
     =& \frac{\alpha}{3\sqrt{N}}
    \sum_{i = -N_{\mathrm{h}}}^{N_{\mathrm{h}}} \sum_{j = -N_{\mathrm{h}}}^{N_{\mathrm{h}}} U_iU_jG(\red L \recolor,i,j,-m) \Delta(i+j-m) \notag \\
    &+ \frac{\alpha}{3\sqrt{N}}
    \sum_{i = -N_{\mathrm{h}}}^{N_{\mathrm{h}}} \sum_{j = -N_{\mathrm{h}}}^{N_{\mathrm{h}}} U_iU_j  G(\red L \recolor,i,j,-m) \Delta(i+j-m \pm N). \label{eq3.1.6} 
\end{align}
The second term on the left-hand side (LHS) of Eq.~\eqref{eq3.1.6} corresponds to the linear force, while the first and second terms on the RHS of Eq.~\eqref{eq3.1.6} describe nonlinear forces. 
The first term on the RHS indicates normal processes, whereas the second term indicates umklapp processes.
\red The second term corresponds to the asymmetric part $\Phi_{\mathrm{a}}(\bm{U})$ in Eq.~\eqref{eq2.2.2a}. \recolor
When the Eq.~\eqref{eq3.1.2} is satisfied, the second term on the RHS vanishes.
Therefore, the symmetricity in the nonlinear lattice~\eqref{eq3.1.1} for \red the map~\eqref{eq2.2.1} \recolor corresponds to the vanishing of umklapp processes.
Considering the condition of the symmetry, Eq.~\eqref{eq3.1.6} can be rewritten as
\begin{equation}
    \ddot{U}_m + 4\red \sin^2 \recolor \frac{\pi m}{N} U_m = \frac{\alpha}{3\sqrt{N}}
    \sum_{i = -N_{\mathrm{h}}}^{N_{\mathrm{h}}} \sum_{j = -N_{\mathrm{h}}}^{N_{\mathrm{h}}} U_iU_jG(\red L \recolor,i,j,-m) \Delta(i+j-m). \label{eq3.1.7} 
\end{equation}
Eq.~\eqref{eq3.1.2} provides the condition of the symmetry, and it is a set of algebraic equations for $\red a_l \recolor$. 
If we obtain the set $\red a_l \recolor$ by solving Eq.~\eqref{eq3.1.2}, the symmetric Hamiltonian \eqref{eq3.1.5} can be obtained.
In the following subsections, we show that the nonlinear lattice with cubic nonlinearity~\eqref{eq3.1.1} exhibits symmetry for the set of \(\red a_l \recolor\) obtained by solving Eq.~\eqref{eq3.1.2}.
It is demonstrated that Eq.~\eqref{eq3.1.2} yields a solution under the condition Eq.~\eqref{eq3.1.3} and \(\red L \recolor=N/2\).
First, the case \(N \rightarrow \infty\) is discussed. 
Subsequently, solutions are obtained to Eq.~\eqref{eq3.1.2} for a finite \(N\).
\subsection{Analytical expressions for the set of \(\red a_l \recolor\)}
For Eq.~\eqref{eq3.1.2}, the following theorem holds.
\begin{thm}
\begin{equation}
    \red a_l \recolor = \frac{2 + (-1)^{\red l \recolor}}{\red l \recolor}, ~\ \red l \recolor =1, 2, \ldots, \red L \recolor \label{eq3.2.1}
\end{equation}
satisfies Eq.~\eqref{eq3.1.2} under the condition Eq.~\eqref{eq3.1.3} and \(\red L \recolor=N/2\) at the limit of infinity \(N\) (\(N \rightarrow \infty\)).
\end{thm}
\begin{proof}
    Denoting \(i_0, j_0, k_0, N_0 \in \mathbb{N}\), such that \(i_0 + j_0 + k_0 = N_0, 2\leq i_0 \leq j_0 \leq k_0 \leq N_0/2-1\).
    Furthermore, considering \(\eta \in \mathbb{N}\) such that \(N = 2\eta N_0\). 
    Subsequently, by replacing \(N\) with \(2\eta N_0\), Eq.~\eqref{eq3.1.2} can be expressed as
    \begin{equation}
        \red \sum_{l=1}^{\eta N_0} \recolor \frac{2+(-1)^{\red l \recolor}}{\red l \recolor} \left(
            \sin {\red l \recolor}I  +  \sin {\red l \recolor}J + \sin {\red l \recolor}K 
        \right)= 0,
    \end{equation}
    where \(I = 2i_0\pi/N_0, J = 2j_0\pi/N_0, K = 2k_0\pi/N_0\). \(\eta \rightarrow \infty\) is equivalent to \(N \rightarrow \infty\). 
    Using relations
    \begin{align}
        \red \sum_{l=1}^{\infty} \recolor \frac{\sin x\red l \recolor}{\red l \recolor} &= \frac{\pi - x}{2}, ~\ 0 < x < 2\pi  \label{eq3.2.2} \\
        \red \sum_{l=1}^{\infty} \recolor (-1)^{\red l \recolor} \frac{\sin x\red l \recolor}{\red l \recolor} &= -\frac{x}{2}, ~\ -\pi < x < \pi \label{eq3.2.3}
    \end{align}
    the following conclusions are drawn:
    \begin{align}
        \red\sum_{l=1}^{\infty}\recolor &\frac{2+(-1)^{\red l \recolor}}{\red l \recolor} \left(
            \sin {\red l \recolor}I + \sin {\red l \recolor}J + \sin {\red l \recolor}K
        \right) = 0.
    \end{align}
    Therefore, \(\red a_l \recolor = \left\{2+(-1)^{\red l \recolor}\right\}/{\red l \recolor}\) satisfies Eq.~\eqref{eq3.1.2} under the conditions of Eq.~\eqref{eq3.1.3} and \(\red L \recolor=N/2\) at the limit of infinity \(N\).
\end{proof}
For finite \(N\), \(\red a_l \recolor\) is expressed as follows:
\begin{equation}
	\red a_l \recolor = \frac{\pi(2+(-1)^{\red l \recolor})}{N\tan(\pi {\red l \recolor}/N)}, ~\ \left(\red l \recolor = 1,2,\ldots,\frac{N}{2}-1\right), ~\ a_{N/2} = 0 \label{eq:finiteN}
\end{equation} 
Derivation details of Eq.~\eqref{eq:finiteN} are described in Appendix B.
\begin{figure}[ht]
\centering
\includegraphics[width = 0.75\linewidth]{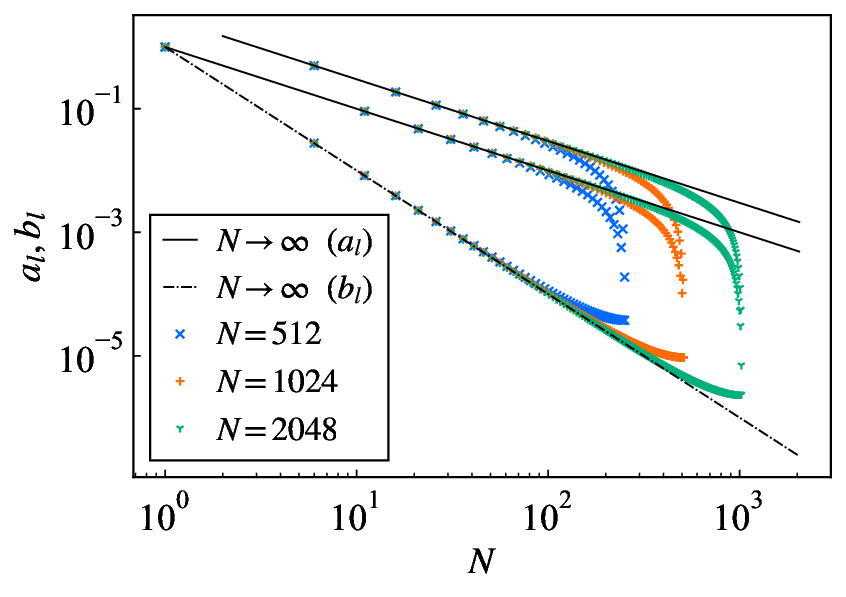}
\caption{Change in \(\red a_l \recolor\) based on Eq.~\eqref{eq:finiteN} for \(N = 512, 1024, 2048\). \(\red b_l \recolor\) is also plotted for comparison. Solid and dashdot lines correspond to the \(\red a_l \recolor\) and \(\red b_l \recolor\) in the limit \(N \rightarrow \infty\), respectively.}
\label{coeff_fig}
\end{figure}
Figure~\ref{coeff_fig} illustrates the change in \(\red a_l \recolor\) according to Eq.~\eqref{eq:finiteN} for \(N =\) 512, 1024, 2048.
The dashed line indicates the value of \(\red a_l \recolor\) for \(N \rightarrow \infty\).
The values of \(\red a_l \recolor\) approach the dashed line asymptotically as \(N\) increases.
Compared with Eq.~\eqref{eq3.2.1}, the drop for larger \(\red l \recolor\) is observed in Eq.~\eqref{eq:finiteN}.
The drop is explained by the finite-size effect.
The coupling constant for a UFL with cubic nonlinearity is given by $\red a_l \recolor \approx {\red l \recolor}^{-1}$. 
The result contrasts with the coupling constant $\red b_l \recolor \approx {\red l \recolor}^{-2}$ for the UFL with quartic nonlinearity, as reported in a previous study~\cite{yoshimura2022}. 
Notably, both $\red a_l \recolor$ and $\red b_l \recolor$ tend to zero as $N\to\infty$, indicating a vanishing coupling constant within the limit. 
However, $\red b_l \recolor$ decreases with an increasing $\red l \recolor$ more rapidly than with $\red a_l \recolor$. 
Therefore, cubic nonlinearity \eqref{eq3.1.1} with solutions \eqref{eq:finiteN} is affected strongly by the effect of truncation.
Particularly, as discussed in Sec.~\ref{sec:level4}, the difference in the coupling constants between \(\red a_l \recolor\) and \(\red b_l \recolor\) affects thermal conductivity for large system sizes. 
\section{\label{sec:level4} Vanishing of umklapp processes in cubic and quartic UFLs}
In this section, we present the numerical results for phonon interactions and energy transport of the proposed UFL. 
In addition to the cubic nonlinearity proposed in Sec.~\ref{sec:level3}, 
quartic nonlinearity is incorporated in our numerical simulations to maintain stability in the UFL.
It should be noted that the quartic nonlinearity is also used in the quartic UFL~\cite{yoshimura2022} and therefore umklapp processes do not occur.
A numerical simulation is performed to determine the phonon modes excited by the perturbation of a single normal mode.
The methodology presented in Ref.~\cite{Raj2019} is used to calculate and characterize these excited modes.
The equation of motion for the UFL is expressed as
\begin{align}
    \ddot{q}_n =& q_{n+1} - 2q_n + q_{n-1} \notag \\
    +& \alpha \red\sum_{l=1}^{N/2}\recolor \frac{\pi\{2 + (-1)^{\red l \recolor}\}}{N\tan \frac{{\red l \recolor}\pi}{N}} \left[
        (q_{n+\red l \recolor} - q_n)^2 - (q_n - q_{n-\red l \recolor})^2
    \right] \notag \\ +& \beta \red \sum_{l=1}^{N/2-1}\recolor \frac{\sin^2 \frac{\pi}{N}}{\sin^2 \frac{{\red l \recolor}\pi}{N}} \left\{
        \left[
            (-1)^{\red l \recolor} q_{n+{\red l \recolor}} - q_n
        \right]^3 - \left[
            q_n - (-1)^{\red l \recolor} q_{n-\red l \recolor}
        \right]^3
    \right\} \notag \\ +& \beta \frac{\sin^2 \frac{\pi}{N}}{2} \left[
        (q_{n+N/2} - q_n)^3 - (q_n - q_{n-N/2})^3
    \right], \label{eq3.4.1}
\end{align}
where the factors \(\alpha\) and \(\beta\) represent the factors of the cubic and quartic nonlinear potentials, respectively.  
The last two terms in the RHS represent quartic nonlinear terms~\cite{yoshimura2022}.
The FPUT-\(\alpha\) lattice with the quartic nonlinearity is employed as the comparison model in which umklapp processes occur:
\begin{align}
    \ddot{q}_n =& q_{n+1} - 2q_n + q_{n-1} \notag \\
    +& \alpha \left[
        (q_{n+1} - q_n)^2 - (q_n - q_{n-1})^2
    \right] \notag \\ +& \beta \red \sum_{l=1}^{N/2-1} \recolor \frac{\sin^2 \frac{\pi}{N}}{\sin^2 \frac{\red l \recolor\pi}{N}} \left\{
        \left[
            (-1)^{\red l \recolor} q_{n+\red l \recolor} - q_n
        \right]^3 - \left[
            q_n - (-1)^{\red l \recolor} q_{n-\red l \recolor}
        \right]^3
    \right\} \notag \\ +& \beta \frac{\sin^2 \frac{\pi}{N}}{2} \left[
        (q_{n+N/2} - q_n)^3 - (q_n - q_{n-N/2})^3
    \right], \label{eq3.4.2}
\end{align}
Periodic boundary conditions are applied to both models.
In the following simulation, the temporal evolution of the normal mode is investigated through the Fourier transform of the particle displacement $q_n$ with a focus on the excited phonon modes arising from umklapp processes.
Initial conditions are defined as
\begin{align}
  q_n (j) &= \frac{2}{\sqrt{N}} \cos \frac{2j\pi n}{N}, \label{eq3.4.3} \\
  \dot{q}_n (j) &= \frac{2}{\sqrt{N}} \omega_j \sin \frac{2j\pi n}{N}, \label{eq3.4.4}
\end{align}
where factor \(j\) is the wave number which satisfies \(0 \leq j \leq N/2-1, ~\ j \in \mathbb{N}\), \(\omega_j = \sin (\pi j/N)\).
We expect umklapp processes arising from cubic nonlinearity following the procedure: interaction between phonon modes \(U_j\) excited other phonon modes \(U_i, ~\ -N/2+1 \leq i \leq N/2-1, ~\ i \in \mathbb{N}\).
Wave number of the new normal mode \(i\) is defined as \(i = \pm 2j\) (normal processes) or \(i = \pm 2j\mp N\) (umklapp processes).
From the definition of \(i\), umklapp processes only occur for \(N/4+1 \leq j \leq N/2-1\).
Parameters are defined as \(N =256, ~\ \alpha = 0.05,\) and \(\beta = 0.1\). 
The numerical simulation is performed with changing \(j\) from \(0\) to \(127\).
The simulations are conducted for a short time in time steps of \(1.0\times 10^{-3}\) to avoid interactions between the excited phonon modes.
\begin{figure}[ht]
       \centering
       \includegraphics[width = 0.75\linewidth]{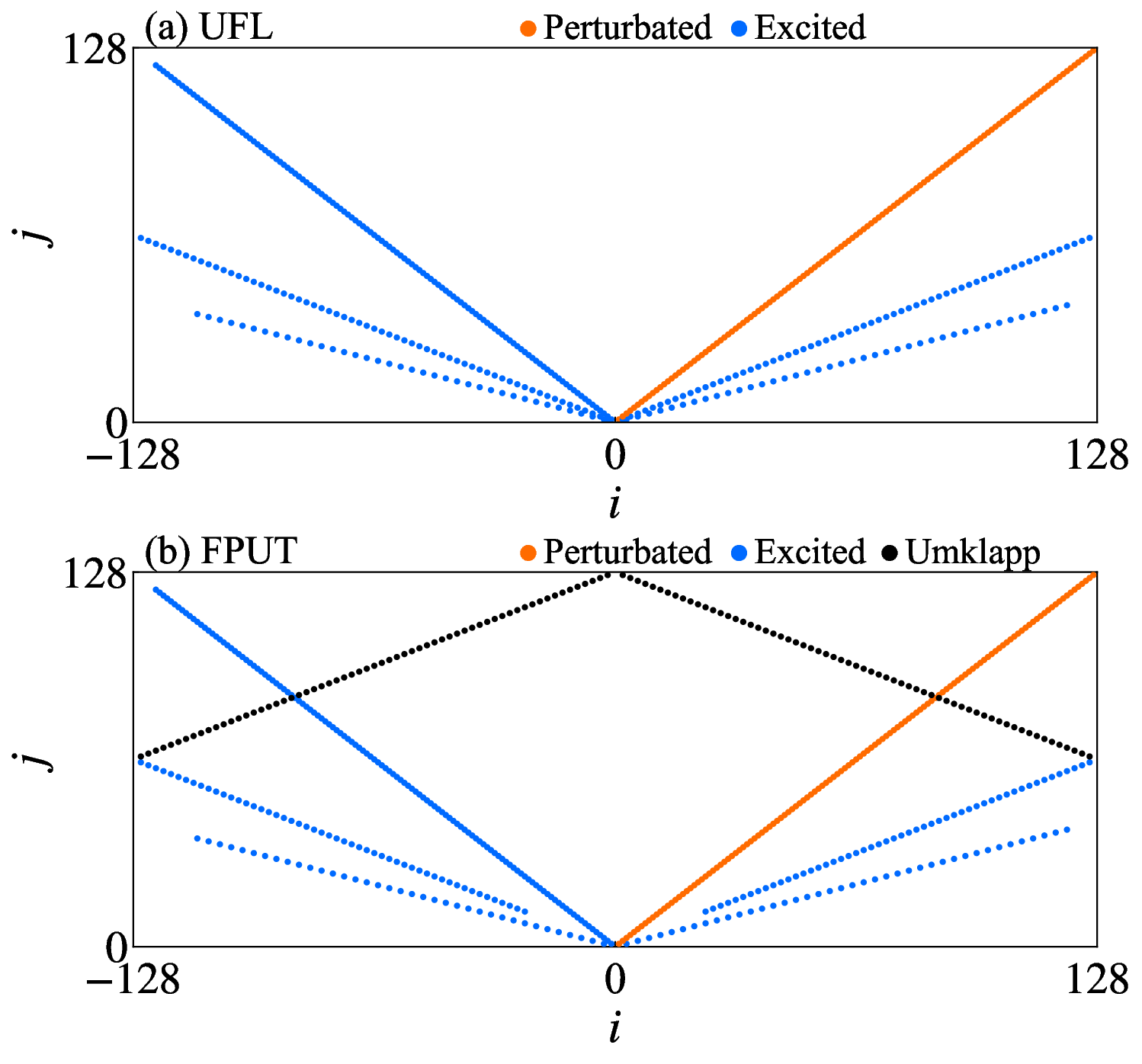}
       \caption{Excited phonon modes (\(i\), \(x\)-axis) from externally perturbed phonon modes (\(j\), \(y\)-axis) on the UFL (top) and the FPUT lattice (bottom). Red, blue, and black dots are separated by the energy spectrum threshold of phonon modes: energy spectra of red dots are greater than \(10^{-3}\), and that of blue and black dots are less than \(10^{-3}\).}
       \label{eps1}
\end{figure}
Figure~\ref{eps1} shows the excitations of the phonon modes caused by the perturbation of a single mode, as described by Eqs.~\eqref{eq3.4.3} and \eqref{eq3.4.4}.
Red dots correspond to the energy of externally perturbed modes.
Blue and black dots show the energy of excited modes.
Especially, black ones shown in Fig.~\ref{eps1}(b) correspond to the energy of phonon modes arising from the three phonon umklapp processes, respectively.
In Fig.~\ref{eps1}(a), the excited phonon modes arising from the umklapp processes are not observed.
Numerical observation verifies the absence of umklapp processes in the UFL.
\section{Ballistic energy transport in cubic and quartic UFLs}
The mechanism of energy transport in a 1D lattice has been discussed based on the scaling law~\cite{lepri1997} between thermal conductivity and the size of systems.
Energy transport in materials is known to follow Fourier's law, represented as \(J = -\kappa\nabla T\), where \(J\) is the energy flux, \(\nabla T\) is the temperature gradient, and \(\kappa\) is the thermal conductivity.
Thermal conductivity is believed to be an intrinsic property of a practical material, and it does not depend on the size of the material at a macroscopic scale. 
\red Conversely, except the rotor model~\cite{PhysRevLett.84.2144,PhysRevLett.84.2381,Iubini_2016}, the anomalous energy transport is observed in 1D nonlinear lattices. \recolor
%
Theoretical~\cite{lepri2003,lepri2023_rev} and experimental~\cite{lepri1997,chang_cnt2008,lee_cnt2017} studies have shown the size dependence of thermal conductivity in 1D lattices. 
The following relation often used to characterize the system is expressed as
\begin{equation}
    \kappa \propto N^{c}, \label{eq4.1.0}
\end{equation}
where \(N\) denotes system size; in 1D lattices, it refers to the number of particles. 
\(0 \leq c \leq 1\) represents the degree of size dependence. 
The case \(c=1\) corresponds to ballistic energy transport.
In ballistic energy transport, the energy flux is independent of \(N\). 
Thus, phonons as carriers of energy, propagate from end to end in the lattices without losing energy. 
A linear lattice is an example of ballistic energy transport because no phonon interactions take place. 
In nonlinear lattices, \(c\) asymptotically approaches \(0\) for larger \(N\).
However, classifying the thermal conductivity in nonlinear lattices using Eq.~\eqref{eq4.1.0} is important.
The thermal conductivity of the proposed UFL is numerically investigated. 
\begin{figure}[ht]
\centering
\includegraphics[width = 0.75\linewidth]{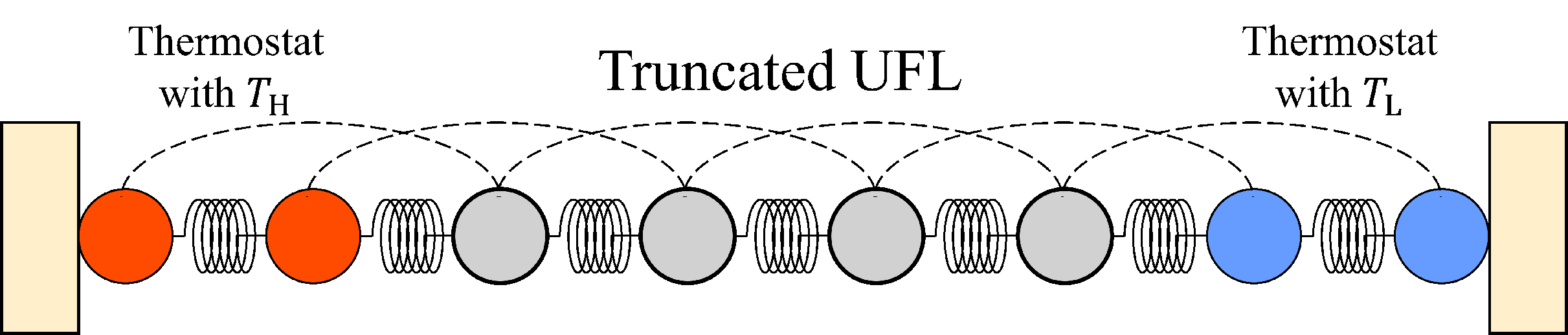}
\caption{Illustration of numerical simulation model. Black dotted lines correspond to long-range interactions. Long-range interactions extend to thermostats. \red Fixed boundary conditions are applied to the left ends in the high temperature thermostat, and the right end in the low temperature thermostat, respectively.\recolor}
\label{fig:simmodel}
\end{figure}
The numerical simulation model is shown in Fig.~\ref{fig:simmodel}.
In the simulation, a finite-size truncated UFL with \(N\) particles is placed in the center. 
Langevin thermostats are connected to both ends of the lattice \red with fixed boundary conditions. \recolor
The equation of motion of the model is derived as
\begin{align}
  \ddot{q}_n =& q_{n+1} - 2q_n + q_{n-1} \notag \\
  +& \alpha \red \sum_{l=1}^{L_3} \recolor \frac{2 + (-1)^{\red l \recolor}}{\red l \recolor} \left[
      (q_{n+\red l \recolor} - q_n)^2 - (q_n - q_{n-\red l \recolor})^2
  \right] \notag \\ +& \beta \red \sum_{l=1}^{L_4} \recolor \frac{1}{\red l^2 \recolor} \left\{
      [(-1)^{\red l \recolor} q_{n+\red l \recolor} - q_n]^3 - [q_n - (-1)^{\red l \recolor} q_{n-\red l \recolor}]^3 
  \right\} \notag \\
  -&\gamma \dot{q}_n + \zeta_n(t) \label{eq4.1.1}
\end{align}
for \(n \in I_{\mathrm{H}} \cup I_{\mathrm{L}}\) and 
\begin{align}
    \ddot{q}_n =& q_{n+1} - 2q_n + q_{n-1} \notag \\
    +& \alpha \red \sum_{l=1}^{L_3} \recolor \frac{2 + (-1)^{\red l \recolor}}{\red l \recolor} \left[
        (q_{n+\red l \recolor} - q_n)^2 - (q_n - q_{n-\red l \recolor})^2
    \right] \notag \\ +& \beta \red \sum_{l=1}^{L_4} \recolor \frac{1}{\red l^2 \recolor} \left\{
        [(-1)^{\red l \recolor} q_{n+\red l \recolor} - q_n]^3 - [q_n - (-1)^{\red l \recolor} q_{n-\red l \recolor}]^3
    \right\} \label{eq4.1.2}
\end{align}
for \(n \in I\), where \(I_{\mathrm{H}} = \{1,2,\ldots,n_0\}\) and \(I_{\mathrm{L}} = \{n_0+N+1, n_0+N+2,\ldots, 2n_0 + N\}\) represent the sets of particle inducers corresponding to the high and low temperature Langevin thermostats, respectively. 
Further, \(I = \{n_0 + 1, n_0 + 2 \ldots, n_0 + N\}\) represents the set of inducers for the truncated UFL.
The parameter \(\gamma\) represents a coefficient.
\(\red L_3 \recolor\) and \(\red L_4 \recolor\) determine truncation length of cubic and quartic long-range interactions, respectively. 
The extensive coupling \cite{iubini2018} is employed as the connection at the boundary between the thermostats and the center part. 
\(n_0\) is defined as
\begin{empheq}[left={n_0=\empheqlbrace}]{alignat=2}
  N/16, & \quad (N \leq 16\red L_3 \recolor) \\
  \red L_3 \recolor. & \quad (N > 16\red L_3 \recolor)
\end{empheq}
The term \(-\gamma \dot{q}_n + \zeta_n(t)\) in Eq.~\eqref{eq4.1.1} represents the Langevin thermostat, where \(\gamma > 0\) is a constant,  $t > 0$ is time,  and \(\zeta_n(t)\) denotes white Gaussian noise. 
\(\zeta_n(t)\) exhibits the following properties:
\begin{align}
    \langle \zeta_n(t) \rangle &= 0, \\
    \langle \zeta_n(t) \zeta_m(s) \rangle &= 2\gamma T \delta_{n,m} \delta(t - s), 
\end{align} 
where  $s > 0$ is time , \(\langle \cdot \rangle\) represents the operation that averages the over-realization of \(\zeta_n(t)\), \(\delta_{n,m}\) is the Kronecker delta, and \(\delta\) is the Dirac delta function.
\(T\) represents thermostat temperature. 
The temperatures \(T_{\mathrm{H}}\) and \(T_{\mathrm{L}}\) are set as high and low temperature sides, respectively. 
The truncated UFL is connected to the thermostat at the \((n_0 + 1)\)-st and \((n_0 + N)\)-th particles,
and there are long-range interactions between the truncated UFL and two thermostats.
Based on these connections, the energy flux \(J_1\) 
which represents the energy transported from the thermostat with \(T_{\mathrm{H}}\) to the \((n_0 + 1)\)-st one per unit time, is expressed as
\begin{equation}
    J_1 = - \left<
        \dot{q}_{n_0+1}\sum_{i \in I_{\mathrm{H}}} K_{i, n_0+1}
    \right>_{\tau},
\end{equation}
where \(\langle \cdot \rangle_{\tau}\) represents an operation that averages over a long period \(\tau\) for any physical quantity \(X(t)\), \(\langle X \rangle_{\tau} = \tau^{-1} \int_0^{\tau} X(t) \mathrm{d} t\). 
\(K_{i,j}\) represents the force acting from site \(i\) to site \(j\).
Similarly, the energy flux \(J_2\),
which represents the energy transported from the \((n_0 + N)\)-th particle to the thermostat with \(T_{\mathrm{L}}\) one per unit time, is calculated as
\begin{equation}
    J_2 = - \left<
        \dot{q}_{n_0+N}\sum_{i \in I_{\mathrm{L}}} K_{n_0+N, i}        
    \right>_{\tau}.
\end{equation}
The effective energy flux \(J\) is defined and computed as
\begin{equation}
    J = \frac{1}{2}(J_1 + J_2).
\end{equation}
Thermal conductivity \(\kappa\) is defined as
\begin{equation}
    \kappa = \frac{J}{T_{\mathrm{H}} - T_{\mathrm{L}}} N.
\end{equation}
In this study, we only focus on energy transport by phonons.
Discrete breathers are known to be excited in various nonlinear lattices with strong nonlinearity~\cite{bagchi2021,st1988,flach2008,Yoshimura2015,PhysRevE.111.054102,wang2020}.
In our numerical simulations, the nonlinearity is relatively small and therefore it is expected that the effect of discrete breathers would be small.
Numerical simulations are performed to solve Eq.~\eqref{eq4.1.2}, and the thermal conductivity $\kappa$ is computed for different lattice sizes \(N =\) 200, 400, 1000, 2000, 4000, 10000, 20000, 100000.
\(\kappa\) is evaluated after every \(4.25\times 10^7\) step when the system is considered to reach a non-equilibrium stationary state.
\red
It is also verified that the time average of heat fluxes \(J_1\) and \(J_2\) are the same for \(N = 10^5\) after \(4.25\times 10^7\) steps passes.
\recolor
The velocity Verlet scheme is employed with a time step of $\Delta t = 1.0\times 10^{-2}$, and the parameters are set as \(\alpha = 0.05, \beta = 0.1,\) \((\red L_3 \recolor, \red L_4 \recolor) = (10, 200), (200, 200)\), \(\gamma = 0.2, T_{\mathrm{H}} = 1.2\), and \(T_{\mathrm{L}}=0.8\).
\begin{figure}[ht]
\centering
\includegraphics[width = 0.75\linewidth]{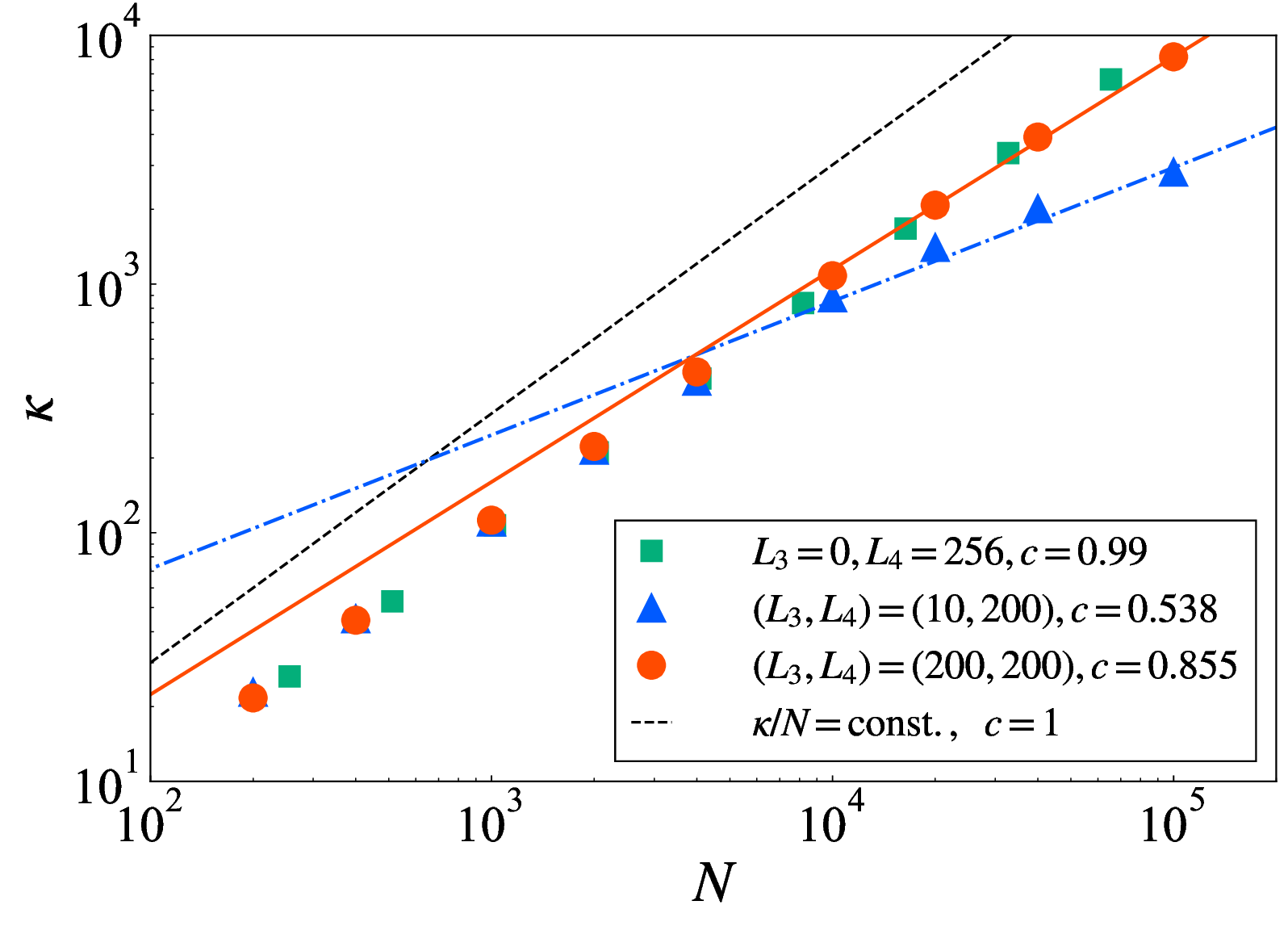}
\caption{Relation between thermal conductivity \(\kappa\) and lattice size \(N\). Dashed line represents the linear case: \(\alpha, \beta = 0\), and shows the ballistic energy transport. 
}
\label{fig1}
\end{figure}
Figure~\ref{fig1} illustrates the relationship between thermal conductivity $\kappa$ and system size $N$. 
The figure presents two sets of numerical results, each corresponding to a different truncation length $\red L_3 \recolor$ for cubic nonlinearity. 
In addition to the obtained results, data for the linear lattice and quartic UFL~\cite{yoshimura2022} are considered for comparison.
When considering the case with \((\red L_3 \recolor, \red L_4 \recolor) = (10, 200)\), shown by the blue triangle, a noticeable reduction in the rate of increase of thermal conductivity $\kappa$ with increasing $N$ is observed, and the energy transport becomes anomalous.  
We obtain a relation \(\kappa \propto N^{0.538}\), shown by blue dashed line.
The \(\kappa\) value is larger than that in the case of the FPUT-\(\alpha\) lattice, \(\kappa \propto N^{0.4}\)~\cite{lepri2000}.
In contrast, for the case \((\red L_3 \recolor, \red L_4 \recolor) = (200, 200)\), \(\kappa\) increases with \(\kappa \propto N^{0.855}\), as shown by the red circles and the red line.
Therefore, thermal transport becomes ballistic asymptotically by increasing \(\red L_3 \recolor\).
Thus, the umklapp processes reduce in the truncated UFL as the truncation length increases.
Consequently, the numerical results verify that the proposed lattice \eqref{eq4.1.2} belongs to the family of UFLs that support ballistic energy transport.
Next, we focus on the difference between the results for the UFL \eqref{eq4.1.2} and the UFL with only quartic nonlinearity, which is indicated by green squares.
Compared with the relation \(\kappa \propto N^{0.855}\), that of the UFL with quartic nonlinearity is near unity within the range of $N\le 10^6$.
The deviation from ballistic transport in the cubic UFL is larger than that in the quartic UFL with the same truncation length. 
As discussed in Sec. \ref{sec:level3}, $\red b_l \recolor$ decreases more rapidly with increasing $r$ compared to $\red a_l \recolor$. 
Consequently, when $r$ is fixed, the effect of truncating the cubic nonlinear potential is more significant than that of truncating the quartic nonlinear interaction.
Another possibility of the deviation can be explained from the viewpoint of energy current of auto-correlation.
Studies have reported that asymmetric potentials such as cubic nonlinearity cause rapid decay of energy current auto-correlation and suggest a diffusive energy transport~\cite{chen2016}. 
Further, more umklapp processes occur in the cubic UFL compared to the quartic UFL owing to the difference in coupling constants of nonlinear interactions.
\red
\begin{figure}[ht]
\centering
\includegraphics[width = 0.75\linewidth]{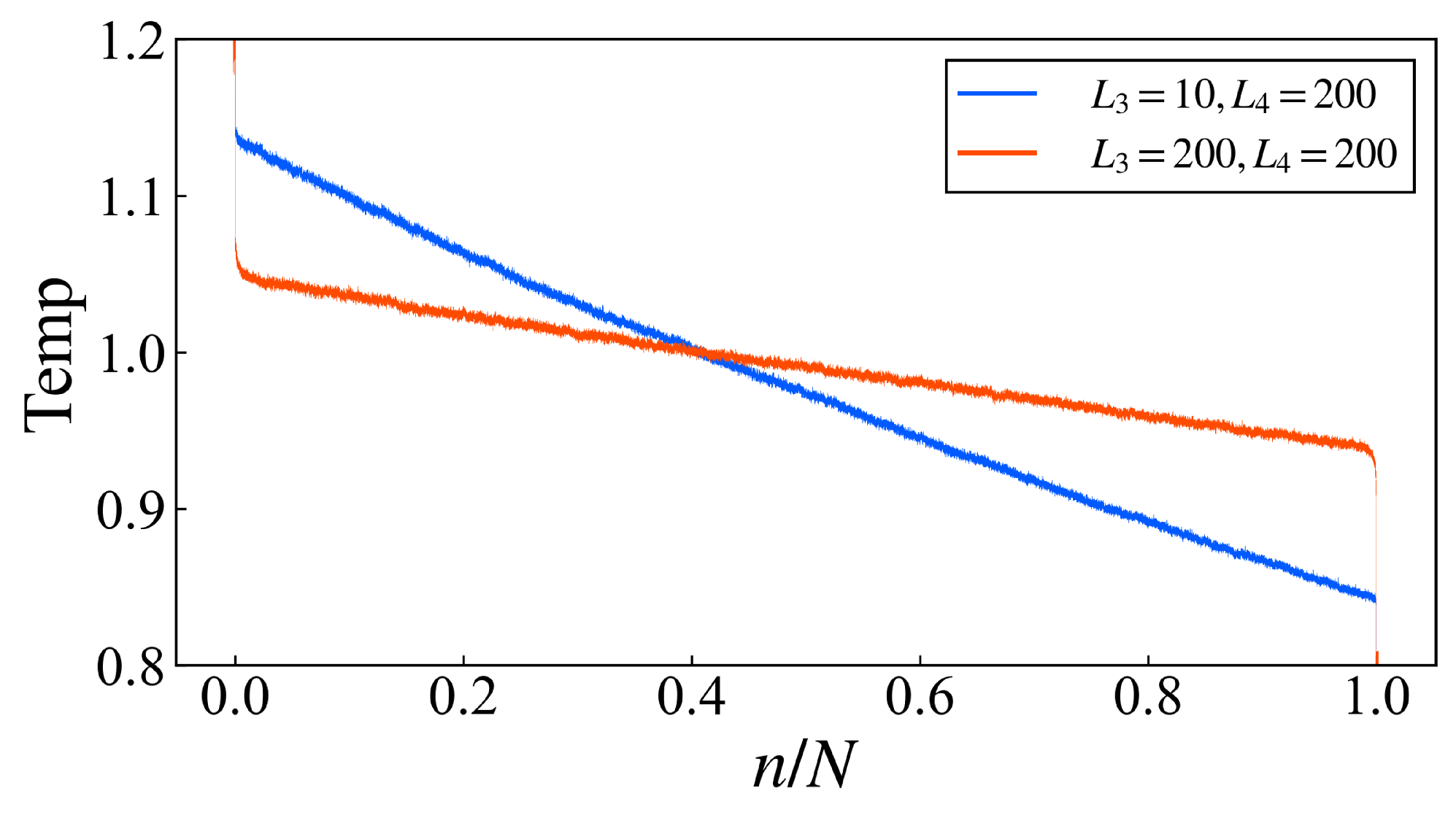}
\red
\caption{Temperature profiles plotted vs \(n/N\) for two cases \((L_3,  L_4) = (10, 200), (200, 200)\). Parameters are \(N = 10^5\), \(T_{\mathrm{H}} = 1.2\), and \(T_{\mathrm{L}} = 0.8\).}
\recolor
\label{fig6}
\end{figure}
The temperature distributions in full systems for two cases \((L_3, L_4) = (10, 200), (200, 200)\) are shown in Fig.~\ref{fig6}, where \(N = 10^5\), \(n\) is the site number of the truncated UFL (\(1 \leq n \leq N\)), and the temperature \(T\) for each particle are defined by the time average of kinetic energy.
It is shown that the temperature gradient away from thermostats close to the flat profile as \(L_3\) becomes larger.
This change of profile of temperatures corresponds to the increase of \(c\).

\recolor
\section{\label{sec:level6}Conclusion}
We constructed the UFL with cubic nonlinearity by considering the symmetry of interparticle potential. 
From the perspective of phonon interactions, umklapp processes arising from cubic nonlinearity are absent in the proposed lattice. 
The cubic nonlinear coupling strength is obtained, and becomes \(\red a_l \recolor = [2+(-1)^{\red l \recolor}]/\red l \recolor\) at the limit of infinity \(N\) (\(N \rightarrow \infty\)).
%
Additionally, we confirmed the absence of umklapp processes by performing numerical simulation for the temporal evolution of phonon modes.
Non-equilibrium simulations were then performed and the thermal conductivity in the truncated UFL was investigated.
The thermal conductivity in the proposed lattice increased from \(\kappa \propto N^{0.538}\) to \(\kappa \propto N^{0.855}\) for varying \(\red L_3 \recolor\). 
Compared to the UFL exhibiting only quartic nonlinearity,  the proposed lattice shows that truncation plays a significant role. 
\section*{Acknowledgments}
The author Y. D. was supported by a Grant-in-Aid for Scientific Research (C), 24K14978, from Japan Society for the Promotion of Science (JSPS).
\appendix
\section{Derivation details of Eq.~\eqref{eq3.1.2}}
Substituting Eq.~\eqref{eq2.1.2} into Eq.~\eqref{eq3.1.1}, we have
\begin{align}
    &H(\bm{U}, \bm{V}, U_{N/2}, V_{N/2}) \notag \\
     &= \frac{1}{2}\sum_{m = -N_{\mathrm{h}}}^{N_{\mathrm{h}}}\left(
        V_mV_{-m} + 4\red \sin^2 \recolor \frac{\pi m}{N} U_m U_{-m}
    \right) + \frac{1}{2}(U_{N/2}^2 + V_{N/2}^2) \notag \\  &+ \frac{\alpha}{3N^{\frac{3}{2}}}
    \sum_{n=1}^{N}
     \sum_{i,j,k = -N_{\mathrm{h}}}^{N_{\mathrm{h}}} {\mathrm{e}}^{-\mathrm{i}\frac{2\pi}{N}(i+j+k)} U_iU_jU_k G(\red L \recolor,i,j,k). \label{eq7.1} 
\end{align}
The summation of \(n\) is \(N\) only when \(i+j+k\) is a multiple of \(N\); otherwise, it is zero.
Since \(-N_{\mathrm{h}} \leq i \leq j \leq k \leq N_{\mathrm{h}}\), considering only three cases is enough; \(i+j+k = 0, \pm N\).
Applying the mapping \(\mathcal{T}_{\lambda}\) to \(U_m, V_m\) in Eq.~\eqref{eq7.1} respectively, we have
\begin{align}
    &H(\mathcal{T}_{\lambda} \bm{U}, \mathcal{T}_{\lambda}\bm{V}, \mathcal{T}_{\lambda}U_{N/2}, \mathcal{T}_{\lambda}V_{N/2}) \notag \\
    &= \frac{1}{2}\sum_{m = -N_{\mathrm{h}}}^{N_{\mathrm{h}}}\left(
        V_mV_{-m} + 4\red \sin^2 \recolor \frac{\pi m}{N} U_m U_{-m}
    \right) \notag \\  &+ \frac{\alpha}{3\sqrt{N}}
     \sum_{i,j,k = -N_{\mathrm{h}}}^{N_{\mathrm{h}}} U_iU_jU_k G(\red L \recolor,i,j,k) \Delta(i+j+k) \notag \\
     &+  \frac{\alpha}{3\sqrt{N}}
     \sum_{i,j,k = -N_{\mathrm{h}}}^{N_{\mathrm{h}}} U_iU_jU_k \mathrm{e}^{\mp \mathrm{i}N\lambda} G(\red L \recolor,i,j,k) \Delta(i+j+k \pm N) \notag \\
     &+ \frac{1}{2}V_{N/2}^2 \mathrm{e}^{-\mathrm{i}N\lambda} + \mathcal{G}(\mathcal{T}_{\lambda}\bm{U}, \mathcal{T}_{\lambda} U_{N/2}). \label{eq7.3} 
\end{align}
To simplify the problem, we assume that \(U_{N/2} \equiv 0\). Under this condition, \(\Phi_{\mathrm{s}}(\bm{V}, \bm{U})\) is defined as the part that is invariant for the mapping \(\mathcal{T}_{\lambda}, \forall \lambda \in \mathbb{R}\), and \(\Phi_{\mathrm{a}}(\bm{U})\) as the rest of the part in Eq.~\eqref{eq7.3}.
Therefore Eq.~\eqref{eq7.3} can be expressed as
\begin{equation}
    H(\mathcal{T}_{\lambda} \bm{U}, \mathcal{T}_{\lambda}\bm{V}) = \Phi_{\mathrm{s}} (\bm{V},\bm{U}) + \Phi_{\mathrm{a}} (\bm{U}), \label{eq7.5}
\end{equation}
where
\begin{align}
    \Phi_{\mathrm{s}} (\bm{V},\bm{U}) &= \frac{1}{2}\sum_{m = -N_{\mathrm{h}}}^{N_{\mathrm{h}}}\left(
        V_mV_{-m} + 4\red \sin^2 \recolor \frac{\pi m}{N} U_m U_{-m}
    \right) \notag \\
    &+  \frac{\alpha}{3\sqrt{N}}
    \sum_{i,j,k = -N_{\mathrm{h}}}^{N_{\mathrm{h}}} U_iU_jU_k G(\red L \recolor,i,j,k) \Delta(i+j+k),  \label{eqA7-6}\\
    \Phi_{\mathrm{a}} (\bm{U}) &= \frac{\alpha}{3\sqrt{N}}
    \sum_{i,j,k = -N_{\mathrm{h}}}^{N_{\mathrm{h}}} U_iU_jU_k \mathrm{e}^{\mp \mathrm{i}N\lambda} G(\red L \recolor,i,j,k) \Delta(i+j+k \pm N). 
\end{align}
\(\Phi_{\mathrm{a}}(\bm{U})\) changes depending on \(\lambda\).
Therefore, if Eq. \eqref{eq7.5} is invariant under the transformation \(T_{\lambda}\), then 
\begin{equation}
    \Phi_{\mathrm{a}}(\bm{U}) \equiv 0 \label{eqA6}
\end{equation}
must hold. 
\(\bm{U}\) in Eq. \eqref{eqA6} depends on \(t\). 
Therefore, the condition is equivalent to
\begin{equation}
    \red \sum_{l=1}^{L} \recolor G(\red L \recolor,i,j,k) = 0, ~\ i+j+k=\pm N. \label{eq7.7}
\end{equation}
When \(i+j+k = \pm N\), \(G(\red L \recolor,i,j,k)\) can be rewritten as 
\begin{equation}
  G(\red L \recolor, i, j, k) = \mp 8\mathrm{i}(-1)^{\red l \recolor} \sin \frac{\pi \red l \recolor i}{N}\sin \frac{\pi \red l \recolor j}{N}\sin \frac{\pi \red l \recolor k}{N}.
\end{equation}
Therefore, we obtain \begin{equation}
    \red \sum_{l=1}^{L} \recolor (-1)^{\red l \recolor} \red a_l \recolor \sin \frac{\pi \red l \recolor i}{N}\sin \frac{\pi \red l \recolor j}{N}\sin \frac{\pi \red l \recolor k}{N} = 0 \label{eq7.8}
\end{equation}
from Eq. \eqref{eq7.7}. Eq.~\eqref{eq7.8} is invariant for the translation \(i, j, k \rightarrow -i, -j, -k\).
Substituting \(-i, -j, -k\) into Eq.~\eqref{eq7.8}, we obtain
\begin{equation}
    -\red \sum_{l=1}^{L} \recolor (-1)^{\red l \recolor} \red a_l \recolor \sin \frac{\pi \red l \recolor i}{N}\sin \frac{\pi \red l \recolor j}{N}\sin \frac{\pi \red l \recolor k}{N} = 0, \label{eq7.9}
\end{equation}
because \(\sin (-\theta) = -\sin \theta, \theta \in \mathbb{R}\),
Herein, Eqs.~\eqref{eq7.8} and \eqref{eq7.9} holds for same \(\red a_l \recolor\). Therefore, we can only consider the case \(i+j+k = N\). 
The ranges of \(i, j, k\) are also restricted as \(2 \leq i \leq j \leq l \leq N_{\mathrm{h}}\) by the inequation
\begin{equation}
    N = i + j + k \leq i + N - 2.
\end{equation}
Finally, the condition of symmetry is expressed as 
\begin{equation}
    \red\sum_{l=1}^{L}\recolor (-1)^{\red l \recolor} \red a_l \recolor \sin \frac{\pi \red l \recolor i}{N}\sin \frac{\pi \red l \recolor j}{N}\sin \frac{\pi \red l \recolor k}{N} = 0  \label{eqA12}
\end{equation}
under the condition
\begin{equation}
    i+j+k=N, ~\ 2 \leq i \leq j \leq k \leq N.
\end{equation}
If Eq.~\eqref{eqA12} holds, then the nonlinear lattice describing the Hamiltonian \eqref{eq3.1.1} is called the UFL.
%

\section{Details of \(\red a_l \recolor\) for finite \(N\)}
Consider the nonlinear term in the equation of motion, given by
\begin{equation}
	\ddot{q}_n = \red \sum_{l=1}^{\infty} \recolor\frac{2+(-1)^{\red l \recolor}}{\red l \recolor}\left[(q_{n+\red l \recolor}-q_{n})^2 - (q_{n}-q_{n-\red l \recolor})^2\right].
	\label{eq:eom}
\end{equation}
The spatial periodic solution is imposed on Eq.~\eqref{eq:eom} to satisfy the relation $q_{n\pm(\red l \recolor+mN)}(t)=q_{n+\red l \recolor}(t)$ for $n, m\in\mathbb{Z}, t\in\mathbb{R}$, where $N$ is a positive even number. 
Substituting the relationship into Eq.~\eqref{eq:eom}, we obtain
\begin{equation}
	\ddot{q}_n = \sum_{m=0}^{\infty}\red \sum_{l=1}^{N} \recolor\frac{2+(-1)^{\red l \recolor +mN}}{\red l \recolor +mN}(q_{n+\red l \recolor}-q_{n})^2
	- \sum_{m=0}^{\infty} \red\sum_{l=1}^{N}\recolor\frac{2+(-1)^{\red l \recolor+mN}}{\red l \recolor+mN}(q_{n}-q_{n-\red l \recolor})^2.
	\label{eq:eom_periodic1}
\end{equation}
As $N$ is even, $(-1)^{mN}=1$ holds and
\begin{equation}
	\ddot{q}_n = \sum_{m=0}^{\infty}\red \sum_{l=1}^{N}\recolor \frac{2+(-1)^{\red l \recolor}}{\red l \recolor+mN}\left[(q_{n+\red l \recolor}-q_{n})^2 - (q_{n}-q_{n-\red l \recolor})^2\right]
	\label{eq:eom_periodic}
\end{equation}
is obtained.
The RHS of Eq.~\eqref{eq:eom_periodic} can be divided into three parts:
\begin{align}
	\ddot{q}_n &= \red \sum_{l=1}^{N/2-1}\recolor \sum_{m=0}^{\infty}\frac{2+(-1)^{\red l \recolor}}{\red l \recolor + mN}\left[
        (q_{n+\red l \recolor}-q_{n})^2 - (q_{n}-q_{n-\red l \recolor})^2
    \right] \notag\\
	&+ \red \sum_{l=N/2+1}^{N} \recolor\sum_{m=0}^{\infty}\frac{2+(-1)^{\red l \recolor}}{\red l \recolor+mN}\left[
        (q_{n+\red l \recolor}-q_{n})^2 - (q_{n}-q_{n-\red l \recolor})^2
    \right] \notag\\
	&+ \sum_{m=0}^{\infty}\frac{2+(-1)^{N/2}}{N/2+mN}\left[
        (q_{n+N/2}-q_{n})^2 - (q_{n}-q_{n-N/2})^2
    \right]. \label{eq:eom_periodic3}
\end{align}
Evidently, the third sum of the RHS vanishes because $(q_{n+N/2}-q_{n})^2 = (q_{n}-q_{n-N/2})^2$ due to the $N$-periodicity of the system.
Additionally, the final component $\red l \recolor =N$ in the second sum of the RHS of Eq.~\eqref{eq:eom_periodic3} vanishes because $q_{n+N}=q_{n-N}=q_{n}$.
Further, we define $\red l' \recolor=N-\red l \recolor$ and substitute the value into the second sum of the RHS of Eq.~\eqref{eq:eom_periodic3} as
 \begin{align}
	\ddot{q}_n &= \red \sum_{l=1}^{N/2-1}\recolor \sum_{m=0}^{\infty}\frac{2+(-1)^{\red l \recolor}}{\red l \recolor+mN}\left[(q_{n+\red l \recolor}-q_{n})^2 - (q_{n}-q_{n-\red l \recolor})^2\right]\notag\\
	&+ \red \sum_{l'=1}^{N/2-1}\recolor \sum_{m=0}^{\infty}\frac{2+(-1)^{N-\red l'\recolor}}{N-\red l'\recolor +mN}\left[(q_{n+N-\red l' \recolor}-q_{n})^2 - (q_{n}-q_{n-N+\red l' \recolor})^2\right].
	\label{eq:eom_periodic6}
\end{align}
Consider the relations $(-1)^{N}=1$, $(-1)^{-\red l'\recolor}=(-1)^{\red l'\recolor}$, $q_{n+N-\red l'\recolor}=q_{n-\red l'\recolor}$, and $q_{n-N+\red l'\recolor}=q_{n+\red l'\recolor}$; then, rewriting $\red l'\to l\recolor$.
By combining the first and second sums in Eq.~\eqref{eq:eom_periodic6}, we have
 \begin{equation}
	\ddot{q}_n = \red \sum_{l=1}^{N/2-1}\recolor \sum_{m=-\infty}^{\infty}\frac{2+(-1)^{\red l \recolor}}{\red l \recolor+mN}\left[(q_{n+\red l \recolor}-q_{n})^2 - (q_{n}-q_{n-\red l \recolor})^2\right].	\label{eq:eom_periodic9}
\end{equation}
Using the formula
\begin{equation}
	\sum_{m=-\infty}^{\infty}\frac{1}{x+m}= \lim_{m\to\infty}\sum_{n=-m}^{m}\frac{1}{x+n}= \pi\cot\pi x,
\end{equation}
we obtain the final result as
\begin{equation}
	\ddot{q}_n = \red \sum_{l=1}^{N/2-1}\recolor \frac{\pi(2+(-1)^{\red l \recolor})}{N\tan\left(
        \pi \red l \recolor/N
    \right)}\left[(q_{n+\red l \recolor}-q_{n})^2 - (q_{n}-q_{n-\red l \recolor})^2\right].	\label{eq:eom_periodic10}
\end{equation}
The result indicates that, in a finite periodic lattice, the coefficient $\red a_l \recolor$ is given by
\begin{equation}
	\red a_l \recolor = \frac{\pi(2+(-1)^{\red l \recolor})}{N\tan(\pi \red l \recolor/N)}, ~\ \red l \recolor=1,2,\ldots,\frac{N}{2}-1. \label{eq:constants}
\end{equation} 
As \(N\) is an even number, \(a_{N/2} = 0\) holds.
%

%
\bibliographystyle{elsarticle-num}

\end{document}